\documentclass[sigconf]{acmart}

\usepackage{url}
\usepackage{booktabs} 
\usepackage{graphicx}
\usepackage{amsmath}
\usepackage{lipsum}
\usepackage[utf8]{inputenc}
\usepackage[english]{babel} 
\usepackage{graphicx}
\usepackage{multirow}
\usepackage{subfig}
\usepackage{tabularx}
\usepackage{tikz}
\newcommand{\tikzcirclered}[2][red,fill=red]{\tikz[baseline=-0.5ex]\draw[#1,radius=#2] (0,0) circle ;}%
\newcommand{\tikzcirclegreen}[2][green,fill=green]{\tikz[baseline=-0.5ex]\draw[#1,radius=#2] (0,0) circle ;}%
\newcommand{\tikzcircleyellow}[2][yellow,fill=yellow]{\tikz[baseline=-0.5ex]\draw[#1,radius=#2] (0,0) circle ;}%
\usepackage{amsmath}
\usepackage{amssymb}
\usepackage{algorithm}
\usepackage[noend]{algpseudocode}
\usepackage{paralist}
\usepackage{color}
\usepackage{footnote}
\makesavenoteenv{tabular}
\makesavenoteenv{table}
\usepackage{stfloats}
\usepackage{courier}
\usepackage[T1]{fontenc}
\usepackage{float}
\usepackage{float}
\floatstyle{plaintop}
\restylefloat{table}
\usepackage{makecell}
\usepackage{amssymb}
\usepackage{pifont}
\usepackage{array}
\newcolumntype{P}[1]{>{\centering\arraybackslash}p{#1}}
\newcolumntype{L}{>{$}l<{$}}
\newcolumntype{C}{>{$}c<{$}}
\newcolumntype{R}{>{$}r<{$}}
\newcommand{\rom}[1]{\uppercase\expandafter{\romannumeral #1\relax}}

\usepackage{siunitx}
\newcommand{\Rey}{\mbox{\textit{Re}}}

\newcommand{\ie}{i.e., }
\newcommand{\eg}{e.g., }

\copyrightyear{2020}
\acmYear{2020}
\acmConference[ICS '20]{2020 International Conference on Supercomputing}{June 29-July 2, 2020}{Barcelona, Spain}
\acmBooktitle{2020 International Conference on Supercomputing (ICS '20), June 29-July 2, 2020, Barcelona, Spain}\acmDOI{10.1145/3392717.3392772}
\acmISBN{978-1-4503-7983-0/20/06}
\setcopyright{rightsretained}

\begin{document}
\title{CFDNet: A deep learning-based accelerator for fluid simulations}
\author{Octavi Obiols-Sales}
\email{oobiols@uci.edu}
\affiliation{
\institution{University of California, Irvine}
}
\author{Abhinav Vishnu}
\email{abhinav.vishnu@amd.com}
\affiliation{
\institution{Advanced Micro Devices, Inc.}
}
\author{Nicholas Malaya}
\email{nicholas.malaya@amd.com}
\affiliation{
\institution{Advanced Micro Devices, Inc.}
}

\author{Aparna Chandramowliswharan}
\email{amowli@uci.edu}
\affiliation{
\institution{University of California, Irvine}
}
\begin{abstract}

  CFD is widely used in physical system design and optimization, where
  it is used to predict engineering quantities of  
  interest, such as the lift on a plane wing or the drag on a motor
  vehicle.
  However, many systems of interest are prohibitively expensive for 
  design optimization, due to the expense of evaluating CFD
  simulations.
  

  To render the computation tractable, reduced-order or 
  {\em surrogate} models are used to accelerate simulations 
  while respecting the convergence constraints provided by the
  higher-fidelity solution. 
  This paper introduces \emph{CFDNet -- a physical simulation and
    deep learning coupled framework}, for accelerating the
  convergence of Reynolds Averaged Navier-Stokes simulations. 
  CFDNet is designed to predict the primary physical properties of the
  fluid including velocity, pressure, and eddy viscosity using a single
  convolutional neural network at its core. 
  We evaluate CFDNet on a variety of use-cases, both extrapolative and interpolative,
  where test geometries are 
  observed/not-observed during training. 
  Our results show that CFDNet meets the convergence constraints of the
  domain-specific physics solver while outperforming it by $1.9 -
  7.4\times$ on both steady laminar and turbulent flows. 
  Moreover, we demonstrate the generalization capacity of CFDNet by testing its prediction on new 
  geometries unseen during training. In this case, the approach meets the CFD convergence criterion while still providing significant speedups over traditional domain-only models. 

\end{abstract}

%
%



\maketitle

\newcommand*\TableQualitative{
	\begin{table*}
		\small
	\begin{center}
		\begin{tabular}{ c  c c c c c c c c c c  }
	 \toprule
	 Related Work & \makecell[l]{Eulerian \\ flows} & \makecell[l]{Laminar \\
      Flows} & \makecell[l]{Turbulent \\ Flows} & \makecell[l]{Viscous\\terms}
      & \makecell[l]{Test\\ Geometry \\ Unseen \\ in Training}
      & \makecell[l]{Predict \\ Mean\\ Velocity } & \makecell[l]{Predict
      \\ Pressure} & \makecell[l]{Predict \\Turbulent \\Viscosity}
      & \makecell[l]{Meets\\ convergence\\constraints} \\ [0.5ex]
 \midrule
			\makecell[l]{\citeauthor{tompson} \cite{tompson}}    & \tikzcirclegreen{1.8pt}  & \tikzcirclered{1.8pt} & \tikzcirclered{1.8pt} & \tikzcirclered{1.8pt} & \tikzcirclegreen{1.8pt} & \tikzcirclered{1.8pt} & \tikzcirclegreen{1.8pt} & \tikzcirclered{1.8pt} &  \tikzcircleyellow{1.8pt}\\
			\makecell[l]{Smart-fluidnet \cite{SC19}}     & \tikzcirclegreen{1.8pt}  & \tikzcirclered{1.8pt} & \tikzcirclered{1.8pt} & \tikzcirclered{1.8pt} & \tikzcirclegreen{1.8pt} & \tikzcirclered{1.8pt} & \tikzcirclegreen{1.8pt} & \tikzcirclered{1.8pt} &  \tikzcircleyellow{1.8pt}\\
			\makecell[l]{\citeauthor{autodesk} \cite{autodesk}} & \tikzcirclegreen{1.8pt}  & \tikzcirclegreen{1.8pt} & \tikzcirclered{1.8pt} & \tikzcirclegreen{1.8pt} & \tikzcirclegreen{1.8pt} & \tikzcirclegreen{1.8pt} & \tikzcirclered{1.8pt} & \tikzcirclered{1.8pt} &  \tikzcirclered{1.8pt}\\
			\makecell[l]{TBNN \cite{ling}}               & \tikzcirclered{1.8pt}  & \tikzcirclered{1.8pt} & \tikzcirclegreen{1.8pt} & \tikzcirclegreen{1.8pt} & \tikzcirclegreen{1.8pt} & \tikzcirclered{1.8pt} & \tikzcirclered{1.8pt} & \tikzcirclegreen{1.8pt} &  \tikzcircleyellow{1.8pt}\\
			\makecell[l]{\citeauthor{maulik} \cite{maulik}}      & \tikzcirclered{1.8pt}  & \tikzcirclered{1.8pt} & \tikzcirclegreen{1.8pt} & \tikzcirclegreen{1.8pt} & \tikzcirclered{1.8pt} & \tikzcirclered{1.8pt} & \tikzcirclered{1.8pt} & \tikzcirclegreen{1.8pt} &  \tikzcirclegreen{1.8pt} \\
			\makecell[l]{\citeauthor{zhu} \cite{zhu}}      & \tikzcirclered{1.8pt}  & \tikzcirclered{1.8pt} & \tikzcirclegreen{1.8pt} & \tikzcirclegreen{1.8pt} & \tikzcirclered{1.8pt}  & \tikzcirclered{1.8pt} & \tikzcirclered{1.8pt} & \tikzcirclegreen{1.8pt} &  \tikzcircleyellow{1.8pt} \\
			\makecell[l]{\citeauthor{unet} \cite{unet}}      & \tikzcirclegreen{1.8pt}  & \tikzcirclegreen{1.8pt} & \tikzcirclegreen{1.8pt} & \tikzcirclegreen{1.8pt} & \tikzcirclered{1.8pt} & \tikzcirclegreen{1.8pt} & \tikzcirclegreen{1.8pt} & \tikzcirclered{1.8pt} &  \tikzcirclered{1.8pt} \\
      \makecell[l]{\bf{CFDNet (This paper)}}            & \tikzcirclegreen{1.8pt}  & \tikzcirclegreen{1.8pt} & \tikzcirclegreen{1.8pt} & \tikzcirclegreen{1.8pt} & \tikzcirclegreen{1.8pt} & \tikzcirclegreen{1.8pt} & \tikzcirclegreen{1.8pt}& \tikzcirclegreen{1.8pt}  &\tikzcirclegreen{1.8pt} \\
 \bottomrule

\end{tabular}
\end{center}
		\caption{\small Comparing CFDNet with SOTA in CFD acceleration using deep learning
    on nine
		features. CFDNet is the first deep learning and physics coupled framework to predict all relevant fluid variables in the entire fluid domain of a turbulent flow to accelerate CFD simulations while meeting the convergence constraints of the original physics solver, whereas prior work \cite{tompson,SC19,ling,zhu} replace specific steps of the original algorithm with a deep neural network (DNN) (\eg find a surrogate of the Poisson solver \cite{tompson,SC19}, estimate the Reynolds stresses \cite{ling}, estimate the eddy viscosity \cite{zhu}) and \cite{autodesk,unet} replace the entire algorithm with a convolutional neural network-based surrogate to estimate the final solution of the fluid variables of interest.}
    \label{tbl:relatedtable}
\end{table*}

}

\section{Introduction}
\label{sec:intro}
\TableQualitative
Computational Fluid Dynamics (CFD) is the \emph{de-facto} method for solving the Navier-Stokes equations -- a challenging multi-scale multi-physics problem.
Typically, the Navier-Stokes equations, a set of Partial Differential Equations (PDEs), are solved iteratively until convergence for each cell of a discretized geometry (also known as \em{grid} or \em{mesh}). 
These simulations are computationally expensive and there are widespread efforts to improve the performance and scalability of solving these systems~\cite{multigrid,salvadore,malaya, Mostafazadeh:aa,behnam}.

The primary advancements in scaling CFD simulations have been achieved through algorithmic innovations and computational optimizations. 
The majority of research has focused on developing {\em domain-specific} optimizations.
At the same time, machine learning (ML) methods -- especially deep learning (DL) algorithms have demonstrated remarkable improvements in a variety of modeling/classification tasks such as computer vision, natural language processing, and high-performance computing~\cite{NIPS2012_4824,43022,liu2016application,Baldi:2014kfa}. 
{\em Our objective in this paper is to explore the potential of DL algorithms to accelerate the convergence of CFD simulations. }

\begin{figure}[b]
\centering
\small
\includegraphics[width=0.9\columnwidth]{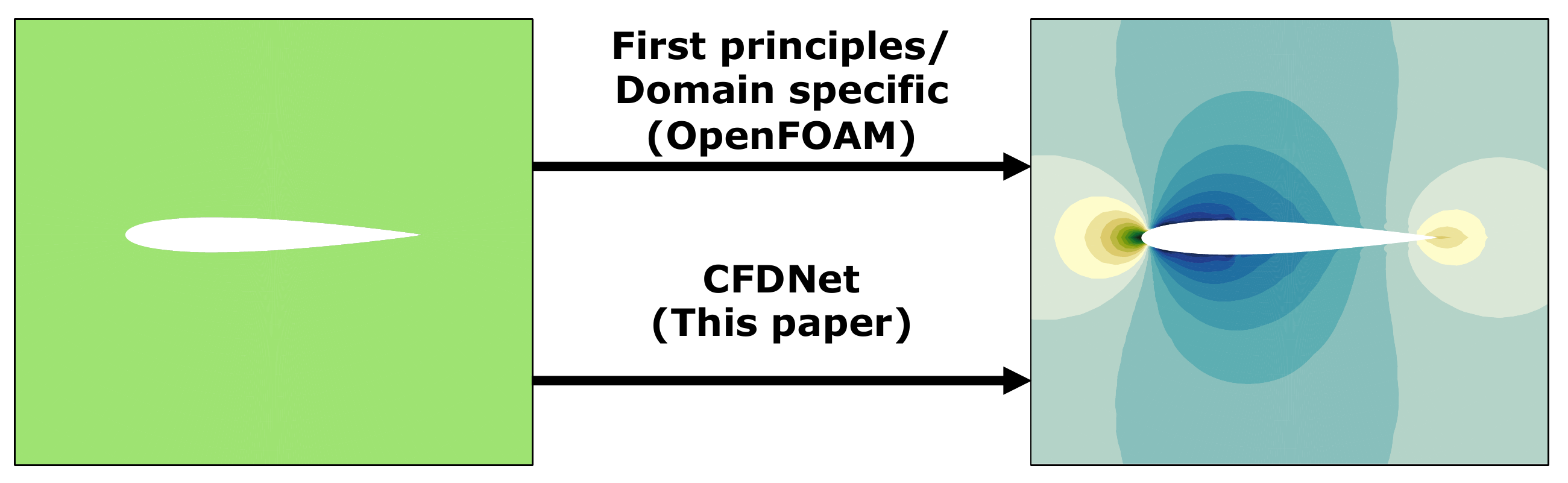}
\caption{\small An example of the input and output of a CFD solver. Our objective with CFDNet is to produce the same output as the physics solver (\ie respect the convergence
constraints) while accelerating the convergence process using DL.}
\label{fig:example-intro}
\end{figure}

Several researchers have applied DL algorithms to accelerate CFD simulations. 
Table~\ref{tbl:relatedtable} summarizes the state-of-the-art (SOTA) methods by comparing them on nine different features. 
While the current approaches have shown the feasibility and potential of deep learning for accelerating fluid flow simulations, they suffer from fundamental limitations. 
First, some of the current approaches do not satisfy the \emph{conservation laws}. 
In particular, DL is used to predict the output variables (such as velocity/pressure) mostly as an end-to-end {\em surrogate}.
This does not guarantee meeting the convergence constraints of the traditional physics-based algorithms which is critical for domain scientists and engineers. 
Second, most of the current methods only predict a partial flow field.
For example, the neural network predicts only a subset of the flow variables which provide incomplete information about the problem.
Others do not support turbulent flows that are common in most industrial applications.
Third, half of the current approaches lack generalization since the test geometry is the same or a subset of the training geometry. 
Generalization using deep learning is challenging especially for unseen geometries.
Overall, we observe that SOTA approaches satisfy up to six out of the nine features shown in Table~\ref{tbl:relatedtable}.

In this paper, we tackle the above limitations and aim to develop new modeling techniques to accelerate CFD simulations, specifically Reynolds Averaged Navier-Stokes (RANS) while respecting the convergence constraints.
We propose CFDNet -- a deep learning framework that combines domain-specific knowledge to meet the same convergence constraints of the physical solver.
An illustration is shown in Figure~\ref{fig:example-intro}.

To address the above challenges, we must consider several choices in designing CFDNet including input/output representation (\eg extracted/raw
features, augmentation), design of the DNN (\eg single/multiple networks for predicted variables, encoder-decoder structure, temporal vs non-temporal output), and algorithm design for combining domain-specific knowledge (\eg warmup before executing DL algorithm and final refinement for adhering to the convergence constraints).  

\textbf{Contributions and Findings.} We evaluate the above design choices and summarize the contributions in this paper as follows.   

\begin{itemize}

	\item We create an input-output representation consisting of multiple channels (one for each predicted flow variable) and a six-layer encoder-decoder convolutional neural network (CNN) with domain-specific activation functions to predict \emph{all} the flow variables of interest (\eg velocity in each direction, pressure, and eddy viscosity of the fluid).
We augment the training set by using each intermediate iteration of the RANS simulation as an input while considering the output of the domain-specific solver as the target variable.

\item We precondition the input to the CNN with a {\em warmup} method, where the first few iterations (determined automatically and adaptively based on the inference use-case) use the domain-specific physics solver before calling the CNN inference. 
We show that the proposed warmup technique to capture domain knowledge reduces the overall refinement steps accelerating the final simulation by $1.9 - 4.6\times$.

\item We propose an \emph{iterative refinement} stage, where the output from the CNN is fed back as the initial condition to the physics solver to meet the domain-specific convergence constraints. 
This circumvents the critical limitation in current approaches where the neural network is modeled as an end-to-end \emph{surrogate}~\cite{autodesk,unet} and there is no guarantee that the conservation laws are satisfied. 

\item We consider several use-cases to evaluate the accuracy, performance, and generalizability of CFDNet as an accelerator.
These include training and prediction on the same geometry (\eg channel flow), training on multiple geometries (ellipses) and predicting on a subset (\eg ellipse with a different aspect ratio) which is common in DL research, and finally, training on multiple geometries (ellipses) and predicting the flow around a new geometry (\eg airfoil or cylinder) which is challenging. CFDNet exhibits remarkable generalizability to geometries unseen during training while achieving $1.9 - 7.4\times$ speedup on laminar and turbulent flows over a widely used physics solver in OpenFOAM. 
\end{itemize}

The rest of the paper is organized as follows. 
We present the necessary CFD background in Section~\ref{sec:background} followed by the CFDNet design in Section~\ref{sec:network}. 
Dataset generation and training recipe are presented in Section~\ref{sec:experiment}, followed by an in-depth discussion of results and analysis in Section~\ref{sec:results}. 
Finally, we present the related work in Section~\ref{sec:related} and conclude in Section~\ref{sec:conclusions}.
\vspace{-0.75em}

\section{Background}
\label{sec:background}
The steady incompressible RANS equations provide an approximate time-averaged solution to the incompressible Navier-Stokes equations. 
They describe turbulent flows as follows: 

\begin{align}
  \frac{\partial \bar{U}_i}{\partial x_i} &= 0\label{eq:continuity}\\ \bar{U}_j
  \frac{\partial \bar{U}_i}{\partial x_j} &= \frac{\partial}{\partial
  x_j}\left[ -\left(\bar{p}\right)\delta_{ij}
  +\left(\nu+\nu_t\right)\left(\frac{\partial \bar{U}_i}{\partial x_j}
  + \frac{\partial \bar{U}_j}{\partial x_i}\right)\right]\label{eq:momentum}
\end{align} 

where $\bar{U}$ is the mean velocity (a 2D or 3D vector field),
$\bar{p}$ is the kinematic mean pressure, $\nu$ is the fluid
viscosity, and $\nu_t$ is the eddy viscosity resulting from Boussinesq's
approximation \cite{boussinesq}. 
Typically, turbulence modeling is used for the eddy
viscosity ($\nu_t$).
The Spalart-Allmaras one-equation model shown below provides
a single transport equation to compute a modified eddy viscosity, $\tilde{\nu}$.

\begin{multline}
\bar{U}_i \frac{\partial \tilde{\nu}}{\partial x_i } = C_{b1} \left(1-f_{t2}\right) \tilde{S} \tilde{\nu} - \left[C_{w1}f_w - \frac{C_{b1}}{\kappa^2} f_{t2} \right] \left(\frac{\tilde{\nu}}{d}^2\right) \\ + \frac{1}{\sigma} \left[\frac{\partial}{\partial x_i} \left( \left(\nu+\tilde{\nu} \right) \frac{\partial \tilde{\nu}}{\partial x_i}\right) +C_{b2} \frac{\partial \tilde{\nu}}{\partial x_j}\frac{\partial \tilde{\nu}}{\partial x_j} \right]
\label{eq:SA}
\end{multline}

Then, we can compute the eddy viscosity from $\tilde{\nu}$ as $\nu_t = \tilde{\nu} f_{v1}$. These equations represent the most commonly-used implementation of the Spalart-Allmaras model.
The terms $f_{v1}$, $\tilde{S}$, and $f_{t2}$ are model-specific and contain, for instance, first order flow features (magnitude of the vorticity). $C_{b1}$, $C_{w1}$, $C_{b2}$, $\kappa$, and $\sigma$ are constants specific to the model, calibrated experimentally. The equations and values of these constants are detailed in \cite{SpalartAllmaras}, the first original reference of the model.

Equations~\eqref{eq:continuity}, \eqref{eq:momentum}, and \eqref{eq:SA} form a system of four PDEs in 2D and five PDEs in 3D. 
We numerically solve the discretized form of these equations on a structured grid with its corresponding boundary conditions (that define the physical boundaries).
The spatially partial derivatives are numerically computed using finite difference methods.
First, the gradients of the flow variables in the grid cell
faces are numerically calculated with a second-order, least-squares interpolation method using the neighboring cells.
Then, for the advection of the velocity and the modified eddy viscosity, we use a second-order, upwind-biased, unbounded scheme. Finally, the diffusion terms in Equation~\eqref{eq:momentum} and Equation~\eqref{eq:SA} are evaluated using Gaussian integration, with a linear interpolation method for the viscosity calculation.
Depending on the non-orthogonality level of the grid, a correction is made to ensure second-order accuracy to compute the surface normal gradient, which is required in the Laplacian calculation.

To find the steady-state solution, we set the necessary boundary conditions and numerically solve the conservation equations until convergence using the semi-implicit method for pressure-linked equations (SIMPLE) algorithm \cite{patankar} outlined in Algorithm~\ref{alg:simple}.
It is an iterative procedure widely used in literature
\cite{SIMPLE-aero,SIMPLE-multigrid,SIMPLE-adaptive} for steady-state problems.
In each iteration, the velocity, pressure, and eddy viscosity fields in the entire grid domain are computed (Lines 2-5) and if the convergence criterion is met in Line~\ref{iteration} (\ie the residual is less than the user-defined tolerance, $\epsilon$), these fields are considered to be final and the \emph{flow has converged} to steady-state.
If not, these fields are intermediate and the algorithm starts the next iteration from Line 2. Algorithm~\ref{alg:simple} is computationally expensive and the goal of this paper is to \emph{short-circuit} the convergence progress by reducing the number of iterations of the RANS simulation.
We will refer to Algorithm~\ref{alg:simple} as the \emph{physics solver} for the rest of this paper.

\newcommand*\SIMPLE{
	\begin{algorithm}[t]
		\small
\caption{\small SIMPLE algorithm.}\label{alg:simple}
\begin{algorithmic}[1]
	\State Guess $\bar{p}^{*}$, an intermediate but incorrect $\bar{p}$ field
	\State Solve $\bar{U}^{*}$, an intermediate but incorrect $\bar{U}$ field, from Equations~\eqref{eq:momentum} \label{init} 
	\State Solve $\bar{p}^{'}$, a correction value for $\bar{p}$, from Equation~\eqref{eq:continuity}
	\State Obtain correct $\bar{U}$ and $\bar{p}$ fields from $\bar{p}^*$, $\bar{U}^*$, and $\bar{p}^{'}$
	\State Solve $\nu_t$ from Equation~\eqref{eq:SA} 
	\If {residual < $\epsilon$} \label{iteration}
	\State \Return done \label{criteriamet}
	\Else 
	\State $\bar{p}^{*}\gets \bar{p}$ 
	\State \textbf{goto} \ref{init}
	\EndIf
\end{algorithmic}
\end{algorithm}

}
 \SIMPLE


\newcommand*\TableErrors{
	\begin{table}[htbp]
		\small
	\begin{center}
  \begin{tabular}{@{}lccc@{}}

    \toprule
	  Network  & Metric    & Best & QoI\\
    \midrule
	  \citeauthor{autodesk} \cite{autodesk}& RME & 1.76\% & $\bar{U}$ \\
	  \citeauthor{unet} \cite{unet} & RME & <3\% & $\bar{U}$, $\bar{p}$ \\
	  Smart-fluidnet \cite{SC19}& MAE & \num{9e-3} & $\rho$ \\
	  TBNN \cite{ling} & RMSE & 0.08 & Reynolds Stresses \\
	  \citeauthor{tompson} \cite{tompson}& $L_2$ norm & 0.872 & $\nabla \cdot U$  \\
	  \citeauthor{maulik} \cite{maulik}& $L_1$ norm & <2e-4 & $\nu_t$ \\
	  \textbf{CFDNet (this paper)} & RME & 0\% & $\bar{U}$, $\bar{p}$ and $\tilde{\nu}$ \\
    \bottomrule
  \end{tabular}
		\caption{\small Errors reported in the literature\label{tab:errors}. Different works have considered different Quantities of Interest (QoI). There is no consensus for an acceptable magnitude of error, and the errors reported as acceptable vary by work.  }
	\end{center}
\end{table}
}

\section{CFDNet: A DL Accelerator for fluid simulations}
\label{sec:network}

Our objective is to develop a function $G$ that accelerates the convergence of the physics solver. 
Let \emph{N} represent the number
of iterations required for convergence with the physics solver in
Algorithm~\ref{alg:simple}, and \emph{I} $(0\leq I \leq N-1)$
represent any intermediate iteration of the solver.
We aim to create a map such that 
$x^{N} = G\left(x^{I};\theta_{wl}\right)$ where $x^{N}, x^{I}\in \mathbb{R}^{m \times
n \times z}$ represent the tensors of $z$ flow variables on a $m\times n$ grid domain at convergence and any intermediate iteration respectively. 

We choose CNN as the candidate for $G$ in this paper (where $\theta_{wl}$ are the $w$ weights in layer $l$ that need to be calibrated) due to its ability to map complex functions -- yet remain domain-agnostic.
Furthermore, neural networks have demonstrated their ability to learn non-linear relationships between input and output variables.
Therefore, we expect the CNN to capture the
nonlinear, multi-dimensional relations present in the Navier-Stokes equations through the nonlinear activation functions. 

In this section, we first describe the different choices in the neural network design
including input/output representation and network architecture.
Then, we present the overall CFDNet framework which combines
domain-knowledge with CNN inference to accelerate fluid simulations and finally, present
our discussion on convergence constraints.

\subsection{Input/Output Representation} 
\label{sec:representation}

We begin with a discussion of the input/output representation (\ie how the tensors $x^I$ and $x^N$ are created).
A 2D cartesian and incompressible RANS flow computes $z = 4$ flow variables -- mean velocity's first component ($\bar{u}$), mean velocity's second component ($\bar{v}$), the mean kinematic relative pressure ($\bar{p}$), and the modified eddy viscosity ($\tilde{\nu}$ \footnote{The Spalart-Allmaras turbulence model uses the transport equation to solve the modified eddy viscosity, so it becomes the fourth variable in our input-output representation.}). 
At each iteration, these four flow variables are updated at every cell of the grid.
Harnessing the "image" nature of the computational grid, where each cell can be
interpreted as a pixel, we generate the input-output representation in
Figure~\ref{fig:inputoutput} as follows:

\begin{figure}[htbp]
  \includegraphics[width=0.5\textwidth]{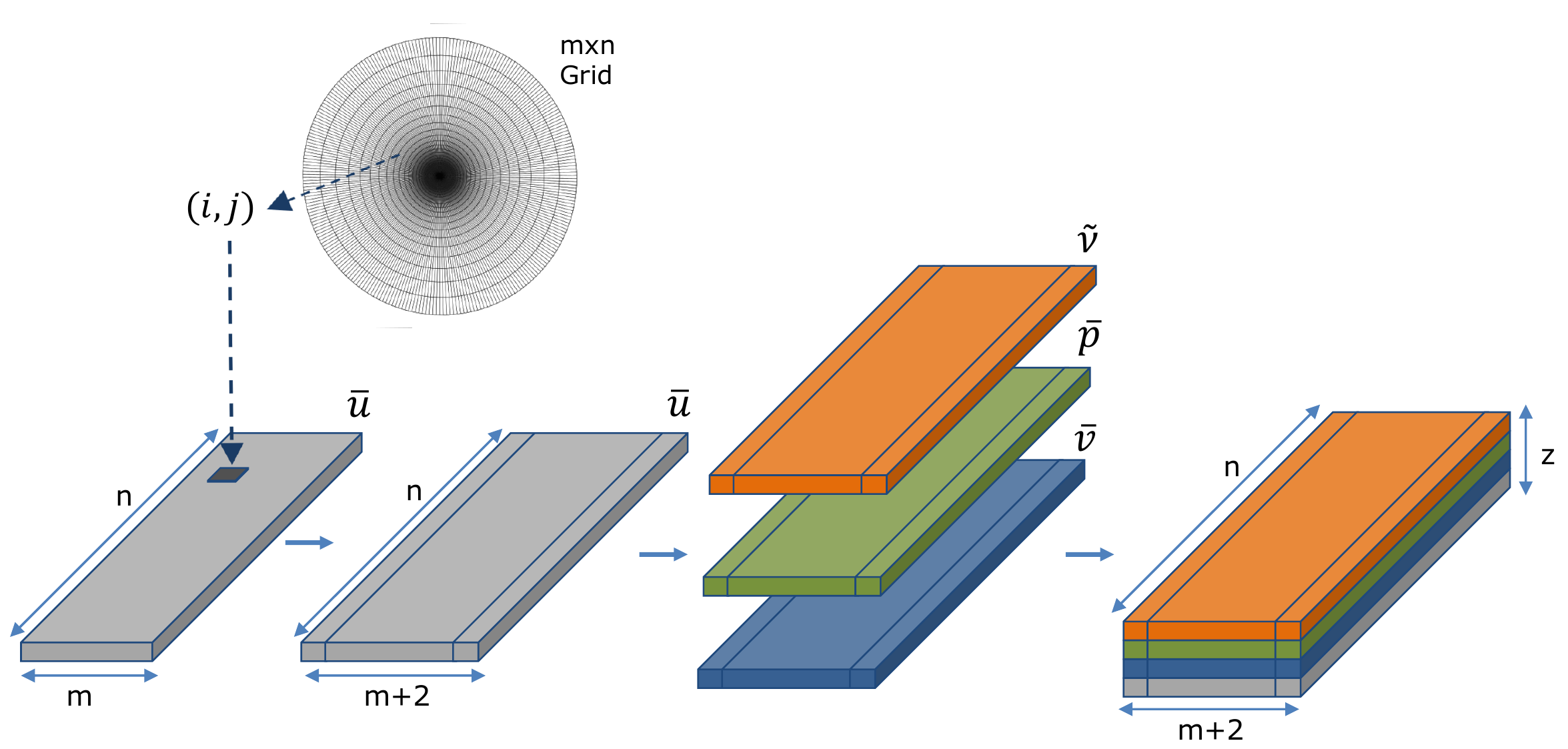}
	\vspace{-1.05em}
  \caption{\small Input-Output Representation in CFDNet. The input and the output
	are images of size $(m+2)\times n \times z$ where $m\times n$ is the
	grid size and $z$ is the number of flow variables (channels). 
	The input image has the grid values of the variables at an intermediate iteration and the output image has the values of the final steady solution.} 
  \label{fig:inputoutput} 
  \end{figure} 

\begin{enumerate}
\item We compose an image of size $m\times n$, where $m \times n$ is the 2D grid size. 
Every pixel $(i,j)$ of this image has the corresponding value of a flow variable, say $\bar{u}$ in the cell $(i,j)$ as shown in the figure. 
The spatial distribution of the pixels needs to respect the logical spatial arrangement of the fluid domain. 
Next, we augment the rows of this image by appending the physical boundary values of the variables from the computational grid (\ie the values at the top and bottom boundaries) leading to an image size of $(m+2) \times n$.
\item We repeat the previous step for the other flow variables of interest -- $\bar{v}$, $\bar{p}$ and $\tilde{\nu}$, leading to four images in total, each of size $(m+2)\times n$.
\item We concatenate the previously generated four images to create the final input/output representation to the CNN which is a tensor of size $(m+2)\times n \times 4$ (\ie an image consisting of four channels). 
\end{enumerate}

Note that both input and output images share the same tensor
dimensionality $(m+2)\times n \times z$. 
The only difference is that the input tensor contains fluid values of
an intermediate iteration while the output tensor is the final,
converged, and steady solution. 

A critical step in the generation of the input/output representation is to \emph{non-dimensionalize} the flow variables. 
This consists of dividing each of the variables in every cell by a flow configuration-specific reference value which is $U_r$ for velocity, $p_r$ for pressure, and $\nu_r$ for eddy viscosity. 
This is a common practice in fluid mechanics~\cite{introduction} to reduce the number of free parameters.
We non-dimensionalize the flow variables for two key reasons both aimed at improving the learning task of the network.
(i) The $z$ flow variables that represent different physical quantities can have a vastly different range of
values (for example, from 0 to 1 \si{\meter\per\second} for velocity and \num{1e-6} to \num{1e-4} \si{\meter^2 \per \second} for the modified eddy viscosity).
Non-dimensionalization rescales the values of the variables to the same range across all $z$ fluid variables. 
(ii) It reduces the number of free parameters; if certain non-dimensionless parameters are significantly smaller than others, they are negligible in certain areas of the flow~\cite{introduction}.

\subsection{Network Design and Architecture}

%

\begin{figure*}[htbp] 
\includegraphics[width=\textwidth]{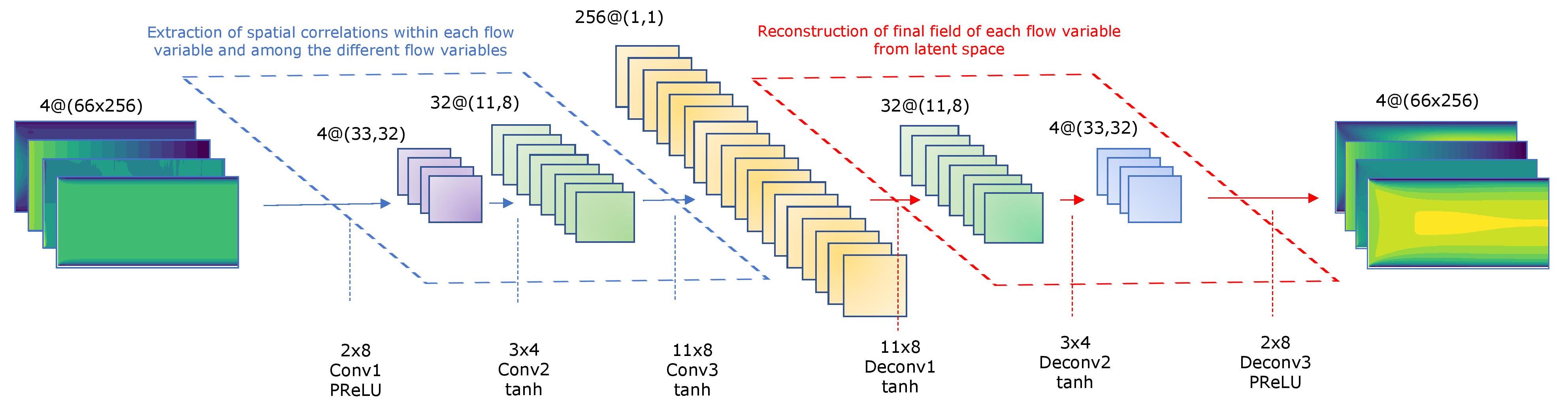}
	\caption{\small CNN architecture. The CNN is a symmetric 6-layer convolution-deconvolution neural network. The first three layers are convolution layers (Conv1 to Conv3) and they are followed by three transposed convolution layers (Deconv1 to Deconv3). The size of each filter is shown at the bottom of the figure along with the activation functions applied to each layer. A striding of the same size as the filter is applied in each layer.}
\label{fig:network}

\end{figure*}

We choose a deep neural network to learn the \emph{surrogate} model for accelerating the RANS equations.
This network is composed of six layers, three convolution and three deconvolution layers. The output is a prediction of \emph{all} the steady-state flow variables ($\bar{u}$, $\bar{v}$, $\bar{p}$, and $\tilde{\nu}$) as shown in Figure \ref{fig:network}. 

The choice of a convolution-deconvolution network is motivated by two main reasons. First, recent works have successfully leveraged similar architectures for physical system emulation \cite{mustafa2019cosmogan,erichson2019physics}. Second, the spatial layout of the flow variables in the input tensor is a result of Equations~\eqref{eq:continuity},~\eqref{eq:momentum}, and~\eqref{eq:SA}. The convolution operator is an optimal candidate to extract the existing spatial correlations in and among the fluid variables (\ie input channels).

We do not include any explicit \emph{domain-specific} feature in the network's input and output because we aim to learn a model that generalizes to a broad number of design use cases comprising of both different geometries - interpolated and extrapolated - and flow conditions. Prior works that embed domain-knowledge in the neural network \emph{learning} task (such as embedding Equation~\eqref{eq:continuity} as the loss function in \cite{tompson}) fail to achieve this generalization limiting its scope. 

The first three convolution layers reduce the dimensionality extracting an abstract representation of the input to encapsulate it in the latent space (middle layer colored yellow). 
From this abstract
representation, which has smaller dimensions than the input, we
reconstruct an output image of the same size as the input. 
The first and last layers use the Parametric Rectified Linear Unit activation function (PReLU \cite{prelu})
to capture any negative value present in the intermediate flow field and to predict the final, real-valued ($\mathbb{R}$) variables in the fluid domain.

\subsection{CFDNet Framework}
\begin{figure*}[htbp]
\centering
\includegraphics[width=\textwidth]{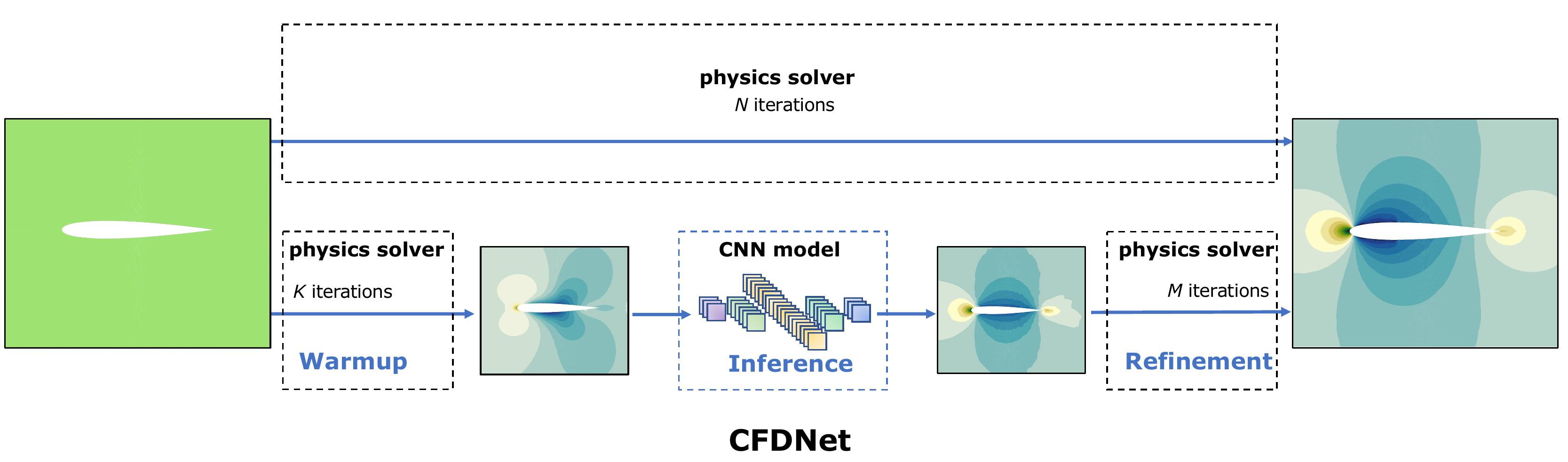}
\caption{\small Comparison of the traditional physics solver simulation with CFDNet. CFDNet integrates the domain-specific physics solver for \emph{warmup}, followed by the neural network for inferring the steady state, and the final iterative \emph{refinement} stage to correct the solution of the CNN and satisfy the convergence constraints. }
\label{fig:workflow}
\end{figure*}

CFDNet is a \emph{physics solver--CNN model--physics solver} coupled framework
instead of a pure surrogate. 
It is designed this way to incorporate domain-specific knowledge in the model inference and to reach the same convergence as the physics solver.
Figure~\ref{fig:workflow} depicts the CFDNet framework and below, we describe the three main stages of the framework and the two key ways we integrate domain-knowledge.

\textbf{\emph{Warmup:}} The first flow field to start the CFD simulation is user-given, therefore it contains no \emph{domain-specific} knowledge. 
So, we let the physics solver carry $K$ initial iterations so that the fluid parameters adapt to the new flow case through Equations~\eqref{eq:continuity},~\eqref{eq:momentum}, and ~\eqref{eq:SA}.
We determine $K$ adaptively without user input by assessing the residual drop from the initial conditions which is an indicator of the evolution of the physics solver.
A residual drop of one order of magnitude is sufficient for the fluid
parameters close to the physical boundaries to capture the geometry and flow
conditions of the new problem. We use this as our warmup stopping criteria.

\textbf{\emph{CNN Inference:}} After warmup, we generate the input image, as described in Section \ref{sec:representation}. 
This input tensor which now has \emph{domain-specific} knowledge of the
new geometry and/or flow configuration is used as an input to the trained CNN model for inference. 
The model predicts the output tensor at steady state.

\textbf{\emph{Iterative Refinement:}} The output of CNN inference may not satisfy known
conservation laws. 
To ensure we meet the convergence constraints of the original algorithm, we feed the output of CNN inference as an input to the physics solver and perform $M$ iterations till convergence. 
The expectation is that the overall number of iterations to reach convergence will be less than the physics solver-only simulation (\ie $K + M < N$) if the data-driven model is successful in \emph{short-circuiting} the simulation.

\subsection{Convergence Criteria and Error of CFDNet}
\label{sec:cfdnetcontribution}
If the CNN inference loss is less than the user-defined error tolerance compared to the ground truth data, CFDNet would return the CNN model's output tensor as the final steady-state flow field (\ie the fluid flow parameters would have reached the convergence criterion defined by the user) bypassing the \emph{refinement} stage.
In this scenario, the CNN model is a \emph{pure} surrogate of the RANS equations with the Spalart-Allmaras one-equation turbulence model. 
However, this approach has significant shortcomings.

First, the above-mentioned convergence criterion is based on error metrics found in the CFD literature~\cite{cfderror}. 
We argue that evaluating the quality of the surrogate through these error
metrics is sub-optimal for two main reasons -- (i) they lack physical meaning, and
(ii) metrics such as $L_1$ norm, mean absolute error (MAE), and root mean
squared error (RMSE) are ill-defined if no information about the order of magnitude of the quantity of interest is provided.
Second, in many approaches, the conservation of mass and momentum is not a constraint in the optimization problem (this is true for our CNN model as well). 
Even though this optimization choice makes the model more generalizable (\eg can be easily extended to support other simulations such as compressible flows), satisfying the conservation laws can be imperative for CFD practitioners/engineers.
Third, ground truth data is typically not available since the goal is to predict the flow field of cases unseen during training. 
Therefore, evaluating the accuracy of the surrogate on error metrics with respect to the ground truth data lacks applicability and predictability in real scenarios.

Alternatively, other works~\cite{SC19,tompson,zhu} have considered finding a DNN-based substitute for only a single step of the iterative algorithm.
For example, finding a surrogate of the Poisson solver in Eulerian flow simulations \cite{SC19}. 
Because this approach modifies the original algorithm, one still needs to evaluate the quality of the result with the error metrics described above, therefore suffering from the same limitations.

We aim to circumvent the above limitations with the \emph{refinement} stage in CFDNet and define its \emph{dual convergence criteria} as follows.
\begin{enumerate}
\item In a traditional CFD simulation, convergence is typically achieved when the residual of all the variables has dropped 4-5 orders of magnitude~\cite{ling} depending on the CFD practitioner. 
We adhere to the same constraints. 
Because the residual is referenced to the initial condition, and our initial
    condition in the \emph{refinement} stage is a field close to the final
    solution, dropping the residual 4-5 orders of magnitude is sufficient to consider the cases fully converged. 
This is the first convergence criterion of CFDNet. 

\item In addition to checking for residual convergence, it is a common practice among domain scientists/engineers to also ensure that the final solution satisfies conservation properties such as \emph{conservation of mass}. 
We also adhere to this convergence constraint and this is the second convergence criterion of CFDNet. 
\end{enumerate}

When both \emph{residual convergence} and \emph{conservation laws} are satisfied, we claim that CFDNet satisfies the \emph{convergence constraints} of the original physics solver and has a solution with 0\% relative mean error (RME) with respect the physics solver solution.
RME is defined as $\sum_{i=1}^{n_c} \frac{ |\hat{y}_i - y_i |}{|y_i|}$ where $n_c$ is the number of cells, $\hat{y}_i$ is the single cell predicted value of the quantity of interest, and $y_i$ is the single cell physics solver value. 
This metric is range and scale-invariant.
Therefore, it is a better indicator of the quality of a prediction. 

%

\newcommand*\TableDatasets{
	\begin{table*}[t]
		\centering
		\normalsize
		\begin{tabular}{@{}p{8mm}p{16mm}p{14mm}p{14mm}p{20mm}p{16mm}p{20mm}p{12mm}p{20mm}@{}}
    \toprule
			Dataset & Flow Type & Flow Condition & \multicolumn{2}{c}{Train} &  \multicolumn{2}{c}{Validation} &  \multicolumn{2}{c}{Test}\\  
			\cmidrule(lr){4-5}
			\cmidrule(lr){6-7}
			\cmidrule(lr){8-9}
			  &  & & Case & Configuration & Case & Configuration & Case & Configuration\\
			\midrule 
			\multirow{4}{*}{A} & \multirow{4}{16mm}{Wall-bounded Flows} & \multirow{4}{*}{Turbulent}  &  \multirow{4}{16mm}{Channel Flow} & \Rey\ = 4200  & \multirow{4}{16mm}{Channel Flow}  &  \multirow{4}{16mm}{\Rey\ = 13000} & \multirow{4}{16mm}{Channel Flow} & \multirow{2}{16mm}{\Rey\ = 5600}\\
					   &                                     &                             &                                &  6800              &                           &                                 &                               &    \\
					   &                                     &                             &                                &  7500              &                           &                                 &                               &   \multirow{2}{*}{13750}  \\
					   &                                     &                             &                                &  12500              &                           &                                 &                               &    \\

			\midrule
			\multirow{6}{*}{B} & \multirow{6}{16mm}{Flow around a solid body} & \multirow{6}{16mm}{Turbulent, \Rey\ = \num{6e5}}  &  \multirow{6}{20mm}{Ellipse} & \emph{a-to-b} ratio: 0.1  & \multirow{6}{16mm}{Ellipse}  &  \multirow{6}{16mm}{\emph{a-to-b} ratio: 0.65} & \multirow{2}{20mm}{Airfoil} & \\
					   &                                     &                                                         &                              &  0.15                     &                               &                                            &                               &    \\
					   &                                     &                                                         &                              &  0.25                     &                               &                                            & \multirow{2}{12mm}{Cylinder}  &    \\
					   &                                     &                                                         &                              &  0.35                      &                             &                                             &                               &    \\
					   &                                     &                                                         &                              &  0.45                      &                             &                                             & \multirow{2}{12mm}{Ellipse}   & \multirow{2}{16mm}{\emph{a-to-b} ratio: 0.3}    \\
					   &                                     &                                                         &                              &  0.55                       &                             &                                            &                               &    \\
		\midrule
			C & Flow around a solid body & Laminar, \Rey\ = 30  &  \multicolumn{2}{c}{Same as Dataset B}  &  \multicolumn{2}{c}{Same as Dataset B}  & \multicolumn{2}{c}{Same as Dataset B}\\
			
    \bottomrule
  \end{tabular}
		\caption{\small Summary of the datasets \label{tab:datasets} used in this paper. Dataset A has different flow configurations of channel flow. Dataset B has simulations of turbulent flow around several solid bodies. Dataset C has the same solid bodies as in B but at laminar flow conditions.}
\end{table*}
}

\section{Experiment Setup}
\label{sec:experiment}

In this section, we first describe the case studies including both laminar and turbulent flows for evaluating the potential of CFDNet as an accelerator.
Then, we describe the construction of the training dataset for training the CNN model and then outline the training process.

\subsection{Case Studies}
\label{sec:cases}
We consider two types of fluid flows -- wall-bounded flows and external aerodynamics which aim to stress different aspects of CFDNet.
For the former, we keep the geometry constant while changing the flow configurations and for the latter, we keep the flow configuration constant while changing the geometry.

\textbf{\emph{Wall-bounded Flows.} } The first case study is turbulent flow in a 2D channel.
Many efforts have been made to understand wall-bounded turbulent flows, especially in confined geometries (\eg pipe and channel designs).
The study of these flow configurations is relevant for a broad range of \Rey\footnote{The Reynolds Number, or \Rey\ is a dimensionless quantity of significance in fluid mechanics and quantifies the flow conditions of the problem.} numbers, on which the final flow regime strongly depends.
An application of CFDNet is to rapidly explore the design space for different ranges of \Rey.

In channel flow, transitional effects from laminar to turbulent flow
occur around \Rey\ $\approx 3000$ \cite{pope}.
Therefore, \Rey\ = 4000 is the lower bound in our design space to
ensure fully developed turbulence flows.
For \Rey\ $>$15000, a different grid resolution than the chosen one should be used to have accurate CFD solutions.
Hence, we scope the range 4000 $<$ \Rey\ $<$15000 for our design exploration.
We train our CNN model on four different configurations: \Rey\ = 4200, 6800, 7500, and 12500, labeled as the training dataset A.
For testing the interpolation and extrapolation performance of CFDNet, we choose \Rey\ = 5600 and \Rey\ = 13750 respectively.

\textbf{\emph{Flow around solid bodies.}}
Understanding flow around solid bodies has been an extensive field of research because of its relevance in aerodynamics design and industrial applications.
It is common practice in aerodynamics design to explore the flow field around variations of the solid body to end up with one that satisfies the designer's performance requirements.
For this reason, we apply CFDNet to flows around geometries that have not been seen during its training phase, in both laminar and turbulent flow conditions.

\begin{figure}[hptb]
\centering
\small
\includegraphics[width=1.0\columnwidth]{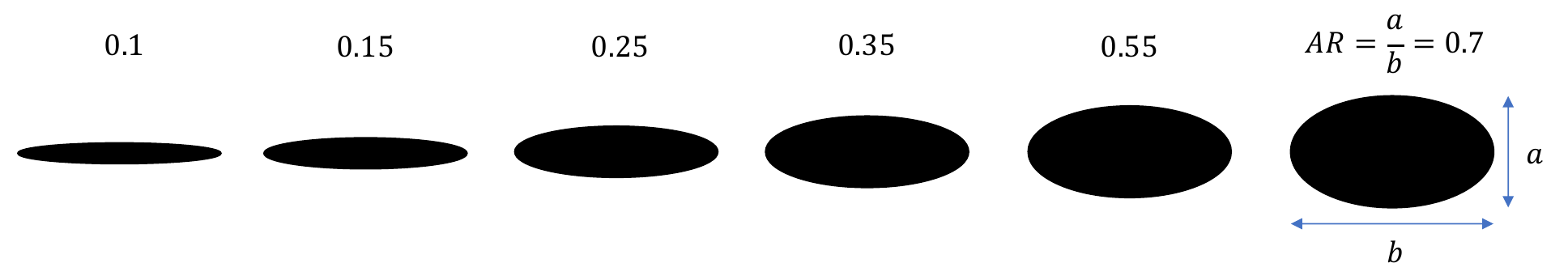}
\caption{\small Ellipse geometries used for training in datasets B and C.}
\label{fig:trainGeometries}
\end{figure}

Training consists of flow around six different ellipses shown in Figure~\ref{fig:trainGeometries}.
The different ellipses are obtained by changing the aspect ratio (\emph{AR}), that is, the ratio of the vertical to the horizontal semiaxis length from $0.1$ to $0.7$.
For the turbulent case, \Rey\ is held constant at \num{6e5} for all the experiments which yield a training dataset labeled B.
Training dataset C has the same geometries as B but at laminar flow conditions (\Rey\ = 30).

\begin{figure}[hptb]
\centering
\small
\includegraphics[width=1.0\columnwidth]{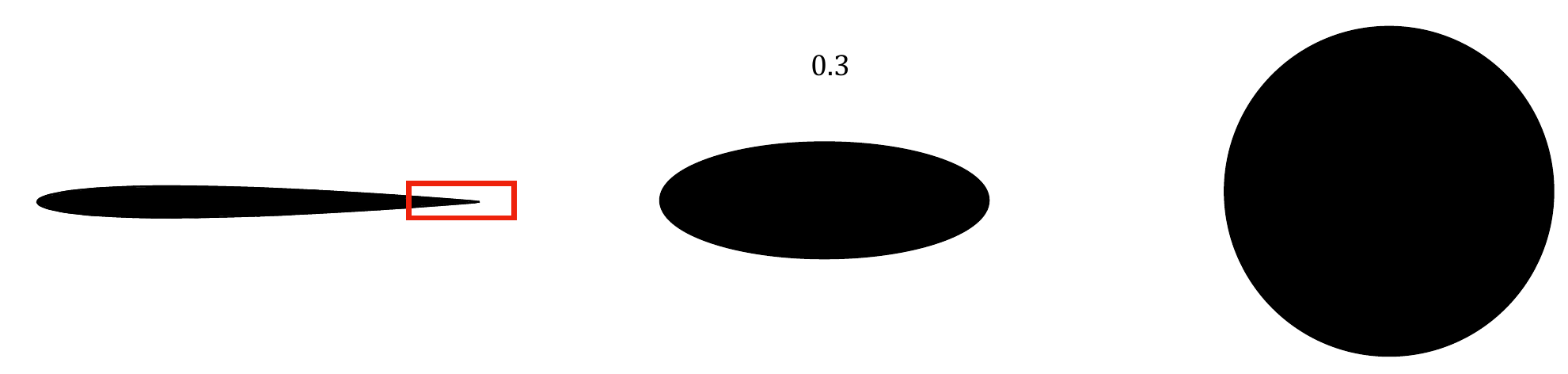}
\caption{\small Airfoil, ellipse, and cylinder used for testing the CFDNet framework. We highlight the trailing edge of the airfoil as a new edge not seen by the network in training.}
\label{fig:testGeometries}
\end{figure}

We test CFDNet on flows around an airfoil, cylinder, and an ellipse (\emph{AR} = $0.3$) not seen in the training phase as shown in Figure~\ref{fig:testGeometries}.

\subsection{Physics Solver in OpenFOAM}
\label{sec:openfoam}
The training dataset is generated by solving the RANS equations together with the Spalart-Allmaras one-equation model \cite{SpalartAllmaras}. The grid is $64\times 256$, a proper grid dimensionality for all cases~\cite{propergrid}.
We use the incompressible solver \texttt{simpleFoam} from OpenFOAM v6 as the \emph{physics solver} in this paper.
The simulations are run till steady state reaching a residual value of \num{1e-6} for the velocity and pressure and \num{1e-4} for the modified eddy viscosity in the channel flow cases.
For the flows around solid bodies, a residual value between \num{1e-5} and \num{1e-6} for the velocity and pressure and \num{1e-4} for the modified eddy viscosity is considered acceptable.
These tolerances are extended practice in CFD simulations \cite{ling}.

Note that our goal in designing the CFDNet accelerator is to use it in tandem with other acceleration techniques commonly used in PDE simulations.
Therefore, we use the \texttt{GAMG} solver for implicitly computing the pressure at every iteration, with a tolerance \num{1e-8} and the \texttt{GaussSeidel} smoother from OpenFOAM.
The velocity and the modified eddy viscosity are computed with the \texttt{smoothSolver} and the \texttt{GaussSeidel} smoother with the same tolerance as the pressure.

\textbf{\emph{Architecture and Libraries.}} All the OpenFOAM simulations are run in parallel on a dual-socket Intel Xeon E5-2630 v3 processor (Haswell-EP).
Each socket has 8 cores, for a total of 16 cores and a theoretical double-precision peak performance of 614.4 GFlop/s.
We use the OpenMPI implementation of MPI integrated in OpenFOAM that is optimized for shared-memory communication.
The grid domain is decomposed into 16 partitions using the OpenFOAM integrated Scotch partitioner and each partition is assigned to 1 MPI process that is pinned to a single core.
The \texttt{numactl -localalloc} flag is set to bind each MPI process to its local memory.

\subsection{Dataset Creation and Preprocessing}
We create three distinct datasets based on the two case studies, as outlined in Section \ref{sec:cases}.
Each dataset has a training set (used for training the CNN model), a validation
set (to ensure the training is not over- or underfitting) and a test set, used
for evaluating the performance of CFDNet.
The recipe for creating the \emph{training} set is as follows:

\begin{enumerate}
\item We perform the OpenFOAM simulation of all training flow configurations.
During the simulation, we take snapshots of the flow field at every intermediate iteration, $I_j$ (for flow configuration $j$) until the flow converges to steady-state.
From each of these iteration snapshots of fluid parameters, we create the tensor image representation of our inputs to CFDNet as detailed in Section ~\ref{sec:representation}.
The resulting number of images is $N_j$, where the first $N_j-1$ are labeled as inputs and the steady-state image at iteration $N_j$ is labeled as the output.
Each of the first $N_j-1$ iterations is independently mapped to steady-state $N_j$ (\ie every iteration of the simulation becomes a training sample whose output is the final steady solution).
Therefore, the training set size for each dataset is $\sum\limits_{j=1}^{n_f} N_j - n_f$, where $n_f$ is the number of flow configurations considered in training (\eg $n_f=4$ for dataset A).
Now, we add the initial conditions (iteration $0$) also as a sample to the dataset and the training set size then becomes $\sum\limits_{j=1}^{n_f} N_j$ for each dataset.

\item We \emph{non-dimensionalize} all the samples. The fluid variables of all images are non-dimensionalized as follows: $\frac{\bar{u}}{U_{r}}$, $\frac{\bar{v}}{U_{r}}$, $\frac{\bar{p}}{U_r^2}$ and $\frac{\tilde{\nu}}{\nu_{r}}$, where subscript $r$ is a reference value. These reference values are flow configuration-specific and characterize each simulation.
\end{enumerate}
The final training set sizes for datasets A, B and C are 6372, 14953, and
12988, respectively.

\subsection{Training}
\label{sec:training}
We train three independent models, one for each dataset A, B, and C.
The CNN (shown in Figure~\ref{fig:network}) is implemented using Keras \cite{keras} and training of the CNN is performed on a Tesla K40c NVIDIA GPU using
the TensorFlow 1.13 backend.
No specific initialization is used in training.
The batch sizes for training are chosen to be 1, 4, and 4 for the datasets A, B, and C respectively. The optimizer is RMSProp and the loss function is mean squared error (MSE).
The learning rate is set to
\num{7e-5} with no decay for all training.
For training set A, 15 epochs were sufficient
to drop the training loss to \num{1.2e-6} and validation loss to
\num{3.2e-6}.
For training set B, after 35 epochs the training loss
reached \num{1.4e-4} and the validation loss \num{2e-4}. For training
set C, the training and validation losses dropped to \num{3.6e-5} and \num{5.1e-5}, respectively after 29 epochs.

The reason for the higher loss in training sets B and C compared to A is because they contain a different set of geometries that present different flow regimes, even though the flow conditions are the same.
Thicker ellipses ($AR=0.55,0.7$) present a significantly more complex, more non-linear flow regime. 
Since the geometry gradients are more accentuated, flow separation from the solid body occurs at these flow conditions. This causes a depression in the rear part of the solid body, leading to negative velocities (recirculation).
In contrast, thinner ellipses ($AR=0.1,0.15$) present a smoother flow behavior. Therefore, the network not only needs to adapt to different grids, but it also needs to adapt to different physical phenomena. In training set A, the geometry and the flow physics are kept the same, therefore the network only has to adapt to a new flow configuration (a new \Rey\ number). The discrepancy between the losses of training set B and C can be explained through the eddy viscosity.
Because of the different flow regimes among the ellipses, the turbulent intensity in the rear part of these solid bodies changes dramatically from ellipse to ellipse (higher in thicker ellipses).
This is a physical phenomenon that does not occur in training set C since there is no turbulence.

\section{Results and Discussion}
\label{sec:results}

Once the CNN is trained and validated, to cover the entire spectrum of its predictive ability, we evaluate CFDNet on three use cases as outlined below.

\textbf{\emph{Observed geometry, different flow conditions (OG-DF).}} We use CFDNet to accelerate the simulation of a flow on a geometry observed during training but on different flow configurations.
Here, our flow case is channel flow and the CNN model is trained with dataset A.
		We test two different flow conditions -- an input velocity to the channel \footnote{Note that referring to \Rey\ number or input velocity to the channel is interchangeable, since \Rey\ $=\frac{U_{i}L}{\nu}$, where $U_{i}$ is the input velocity, $L$ is the channel height (constant) and $\nu$ is the fluid laminar viscosity (constant).}
of $0.56$ \si{\meter\per\second} and $1.375$ \si{\meter\per\second}. With the former, we evaluate CFDNet's capacity to interpolate to a new flow condition, whereas with the latter, we test its ability to extrapolate.

\textbf{\emph{Subset geometry, same flow conditions (SG-SF).}} Test geometry is a \emph{subset} of the training dataset.
Here, we aim to test the ability of the CNN to interpret the edges of the new geometry which can be formed using a linear combination of the training geometries (\ie interpolation).
In this use-case, the flow conditions are kept the same throughout the experiment.
Here, our flow case is external aerodynamics and the CNN models are trained with datasets B and C for turbulent and laminar flows respectively.
The geometries used for training are depicted in Figure \ref{fig:trainGeometries} and the test geometry is also an ellipse (with an unseen-in-training \emph{AR}) shown in Figure~\ref{fig:testGeometries}.

\textbf{\emph{Different geometry, same flow conditions (DG-SF).}} Test geometry is \emph{different} from the training dataset.
Here, we test CFDNet's capacity for generalization where the edges of the test geometries have not been seen previously by the CNN. 
To underscore that, we reuse the CNN models trained for the previous use case SG-SF.
We test the performance of CFDNet on the NACA0012 airfoil \cite{NACA0012}, the geometry of which follows the mathematical definition in \cite{NACA0012geometry} and a cylinder shown in Figure~\ref{fig:testGeometries}.

The accuracy and performance of CFDNet are evaluated by comparing it against the traditional physics solver-only simulation.
We first evaluate the accuracy of our convolutional neural network prediction and the end-to-end CFDNet coupled framework.
Then, we evaluate the performance of CFDNet and finally examine the importance of the \emph{warmup} method in accelerating the convergence of CFDNet, especially for new test geometries.

\subsection{Model and CFDNet Accuracy}
One of the goals of this paper is to accelerate flow simulations using deep learning without relaxing the convergence constraints of the physics solver.
Since the network is trained to predict the final steady-state (iteration $N$), we hypothesize that the inference is "close" to the final solution.
In this section, we test this hypothesis by comparing the CNN model's output with CFDNet's output, i.e.,  compare the solution with and without \emph{refinement}.
At the same time, we also compare both to the physics solver solution qualitatively and quantitatively.

\begin{figure}[htbp]
  \includegraphics[width=\linewidth]{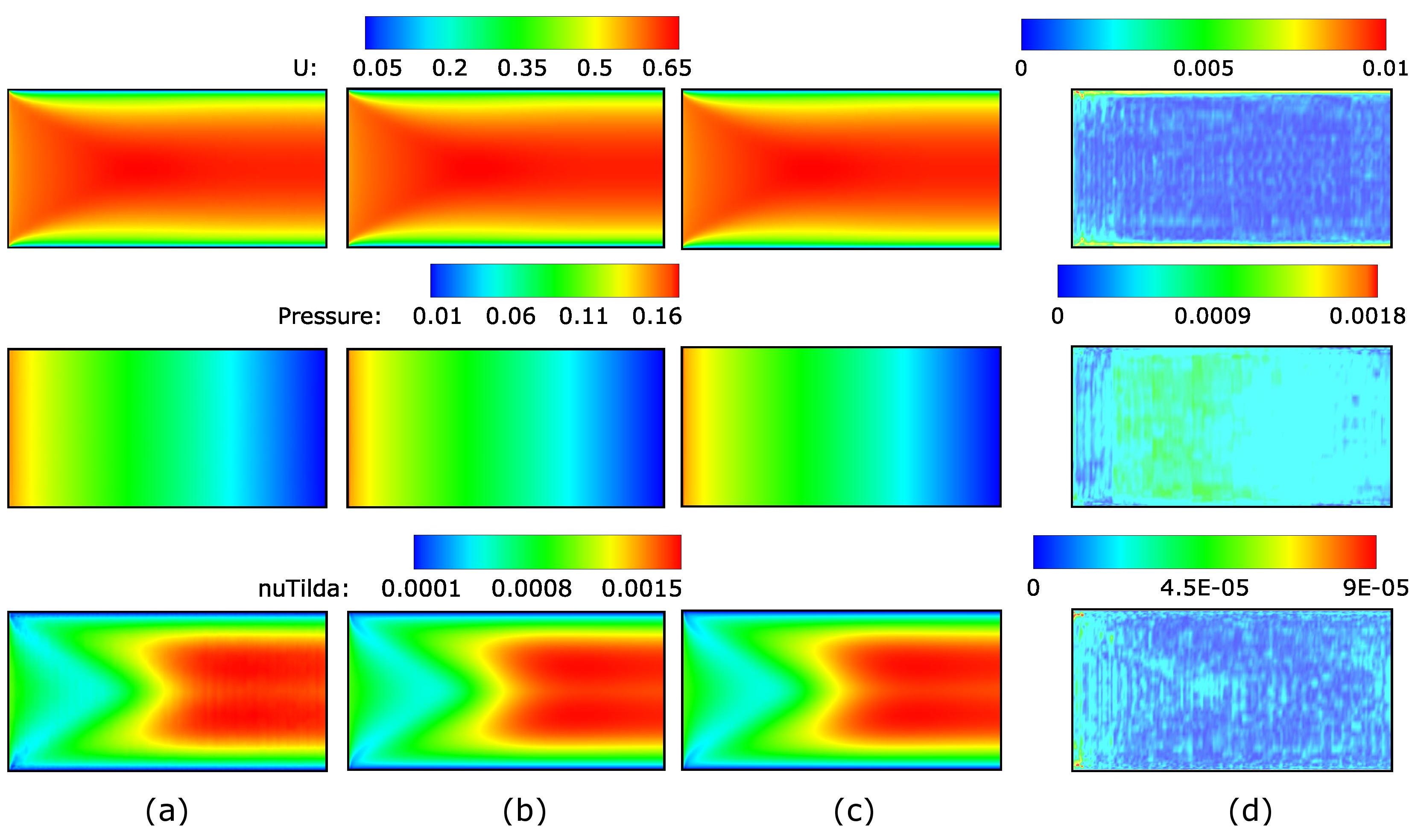}
	\caption{\small Velocity field in \si{\meter\per\second} (top), Kinematic pressure field in \si{\meter^2\per\second^2} (middle), and Modified eddy viscosity field in \si{\meter^2\per\second} (bottom) for channel flow at \Rey\ = $5600$. (a) \emph{warmup} + \emph{inference} (no \emph{refinement}), (b) CFDNet, (c) physics solver in OpenFOAM, and (d) per-cell absolute error between (a) and (c).}
  \label{fig:channelFlow}
\end{figure}

\begin{figure}[htbp]
  \includegraphics[width=\linewidth]{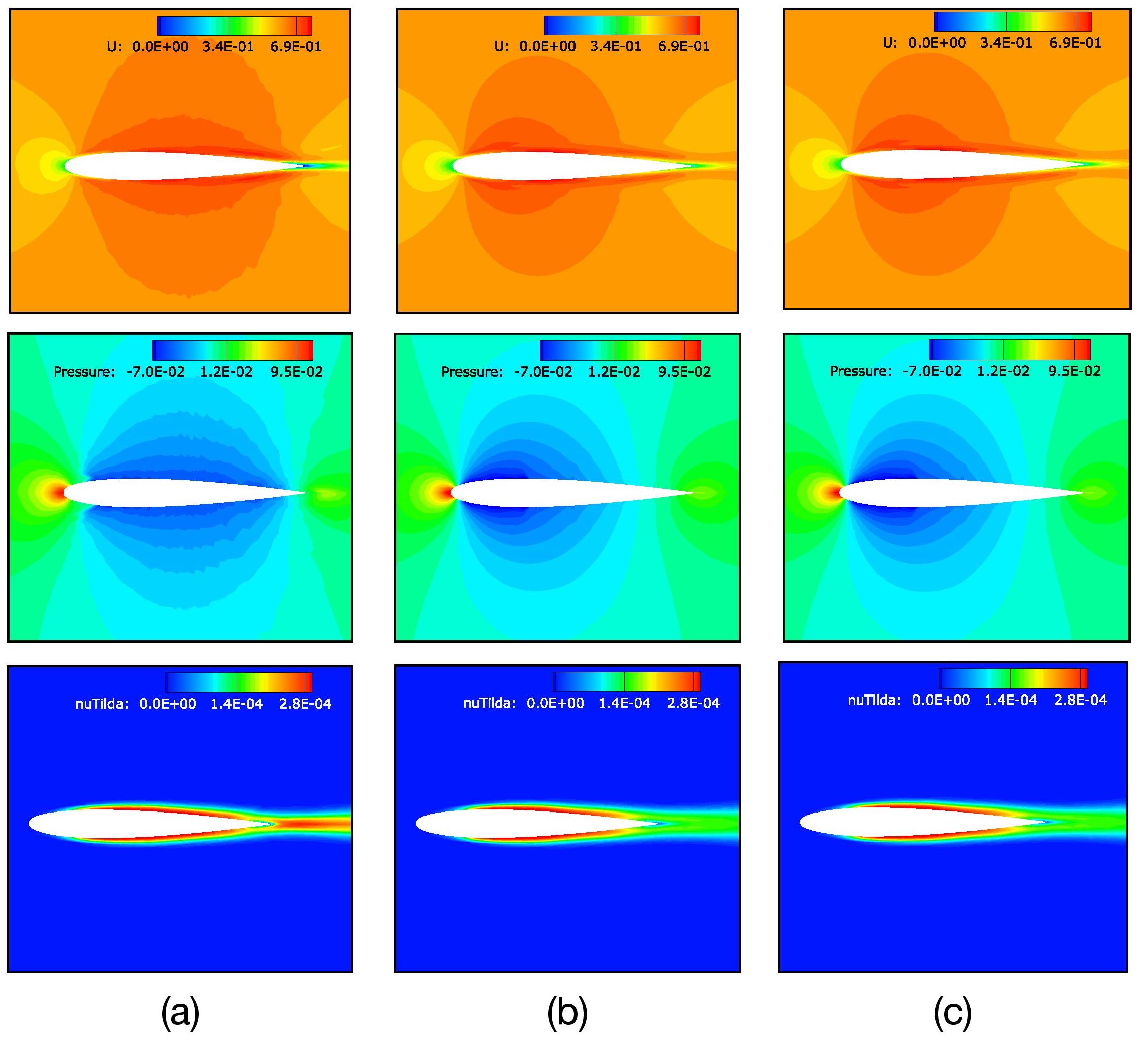}
	
	\caption{\small Velocity field in \si{\meter\per\second} (top), Kinematic pressure field in \si{\meter^2\per\second^2} (middle), and Modified eddy viscosity field in \si{\meter^2\per\second} (bottom) around the airfoil at \Rey\ = \num{6e5}. (a) \emph{warmup} + \emph{inference} (no \emph{refinement}), (b) CFDNet, and (c) physics solver in OpenFOAM. \label{fig:airfoil}}
	
\end{figure}

\begin{figure}[htbp]
  \includegraphics[width=\linewidth]{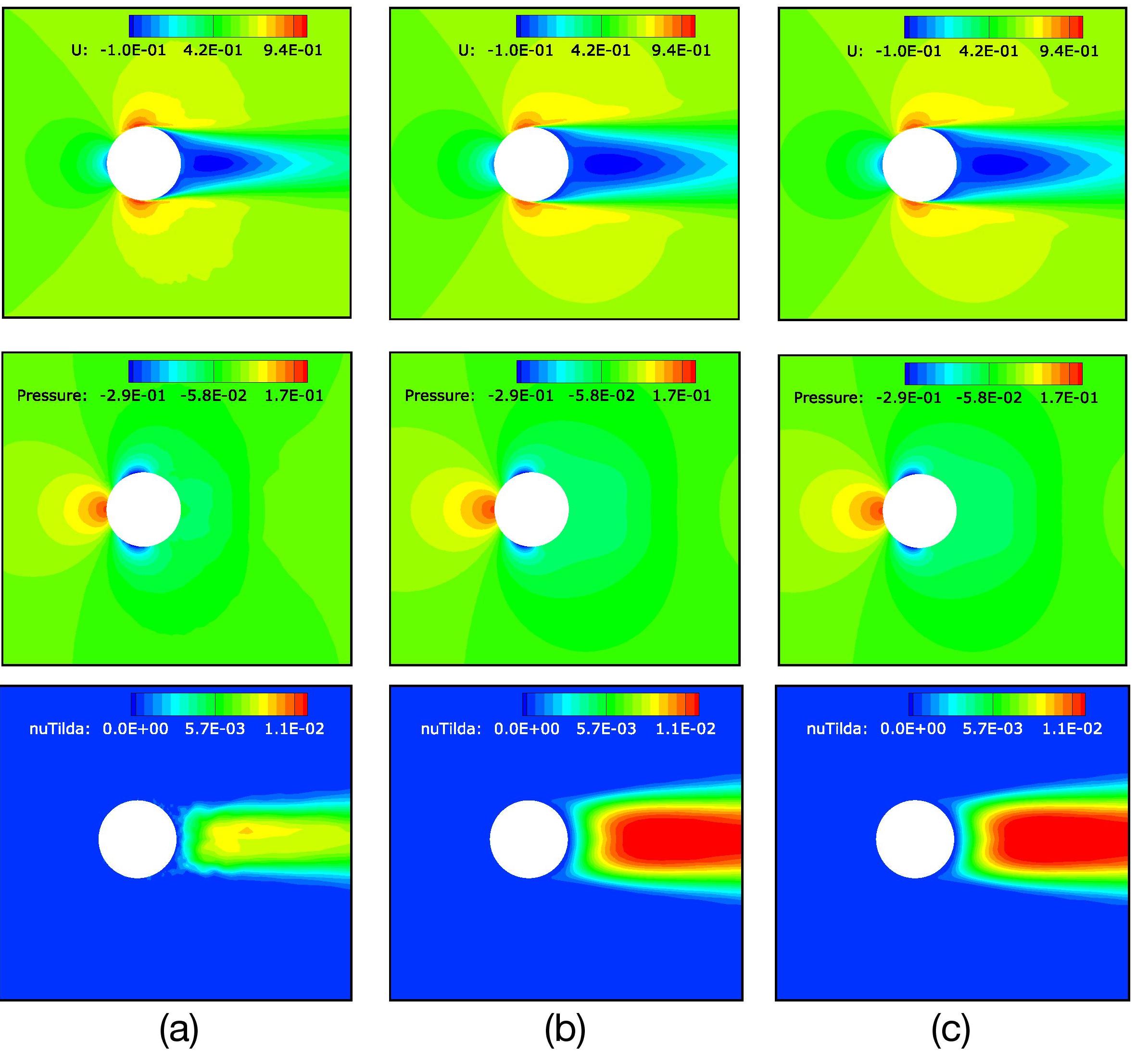}

	\caption{\small Velocity field in \si{\meter\per\second} (top), Kinematic pressure field in \si{\meter^2\per\second^2} (middle), and Modified eddy viscosity field in \si{\meter^2\per\second} (bottom) around the cylinder at \Rey\ = \num{6e5}. (a) \emph{warmup} + \emph{inference} (no \emph{refinement}), (b) CFDNet, and (c) physics solver in OpenFOAM. \label{fig:cylinder}}
\end{figure}

Figure~\ref{fig:channelFlow} shows the qualitative results for the OG-DF interpolated case.
The leftmost column (a) presents the velocity field after \emph{warmup} +\emph{inference} (\ie the CNN model's output).
We can observe the similarity between the output from the CNN and the physics solver solution in the third column (c). However, this is not the case for the DG-SF airfoil at \Rey\ = \num{6e5}. 
We can see that the CNN predicted pressure field in Figure~\ref{fig:airfoil} (a) qualitatively differs from the physics solver solution (c)  mostly in the front part of the airfoil. 
Similiar to the airfoil, we can visually conclude that the prediction of the modified eddy viscosity for the DG-SF cylinder in Figure~\ref{fig:cylinder} with the CNN (a) is qualitatively far from the physics solver (c).
This discrepancy in the quality of the prediction between the OG-DF channel flow case and the DG-SF airfoil/cylinder case is expected from the training results (refer to Section~\ref{sec:training}) and the challenge of each test case individually - interpolate a new flow condition vs. extrapolate to an unseen geometry. 

Quantitatively, the per-cell absolute error between the CNN prediction and the physics solver solution can be seen in Figure~\ref{fig:channelFlow} (d) for OG-DF channel flow case. 
The per-cell absolute error is 2 to 3 orders of magnitude lower than the variable value, yielding a RME less than 2\% for all flow variables, which is in line with acceptable RME reported in the literature, as seen in Table~\ref{tab:errors}. 
This suggest that the CNN (\ie warmup + inference) is a promising approach as a pure surrogate when the geometry is the same (OG-DF) but not otherwise.

\TableErrors

Recall from Section~\ref{sec:cfdnetcontribution} that the quantitative analysis
through the above error metrics alone is insufficient. 
Moreover, Table~\ref{tab:errors} shows that there is no consensus among SOTA approaches on the best error metric to adopt.
Therefore, we conduct the conservation of mass \emph{check}. 
Because the flow is incompressible, the velocity field has to satisfy the conservation of mass Equation~\eqref{eq:continuity}, \ie $\nabla \cdot U = 0$. 
OpenFOAM provides a tool to numerically compute the divergence of the velocity for a user-given flow field. 
Table~\ref{tab:mass} compares the result of the tool for the flow fields from CNN model-only (\emph{warmup} + \emph{inference}), CFDNet (\emph{warmup} + \emph{inference} + \emph{refinement}), and the physics solver-only simulations.

\newcommand*\TableMass{

	\begin{table}[htpb]
		\small
	\begin{center}
		\begin{tabular}{@{}p{30mm}p{18mm}P{10mm}p{10mm}@{}}

    \toprule
	  Test Case  & \emph{warmup}+\emph{inference} only  & CFDNet & physics solver\\
    \midrule
		channel flow \Rey\ = $5600$ & \num{3.12e-4} & <\num{1e-10} & <\num{1e-10} \\
			airfoil \Rey\ =\num{6e5} & \num{8.15e-3} & <\num{1e-10} & <\num{1e-10} \\
			cylinder \Rey\ = \num{6e5} & \num{9.8e-3} & <\num{1e-10} & <\num{1e-10}\\
    \bottomrule
  \end{tabular}
		\vspace{-1.25em}
		\caption{\small Mass local error. The value reported is the average over all the cells. The CNN model does not result in a divergence-free field. This is a major motivation for refinement after the initial warmup and inference steps. After the final refinement, the field is divergence-free. \label{tab:mass}}
	\end{center}
\end{table}

}

\TableMass

Because analytically the divergence of the velocity needs to be exactly 0, numerically it should be close to the machine round-off errors. 
Table~\ref{tab:mass} illustrates how CFDNet and the physics solver yield a divergence-free field (\ie both satisfy conservation of mass), whereas the CNN model's output is far from the expected tolerance. 
Even with a CNN model's prediction - for the channel flow case - yielding a 2\% RME, the \emph{iterative refinement} becomes necessary to satisfy the conservation laws and meet the convergence constraints of the original physics solver.

\subsection{Performance Analysis}

\begin{figure*}[htbp]
\includegraphics[width=\linewidth]{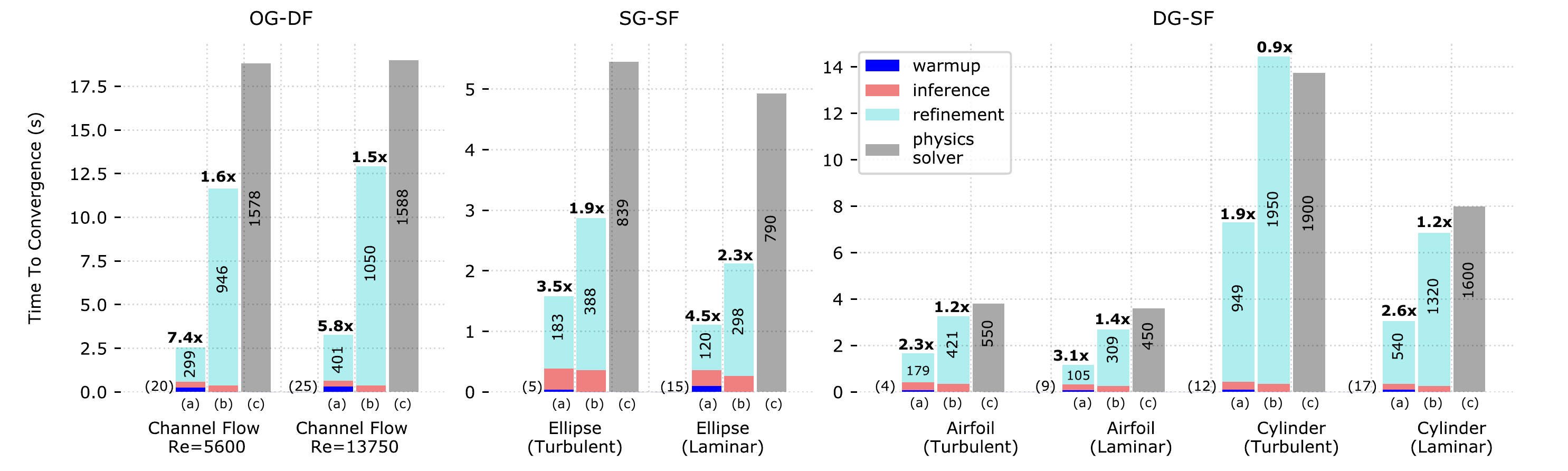}
	\caption{Breakdown of running time to convergence for each test case. Time to convergence of (a) CFDNet w/ \emph{warmup}, (b) CFDNet w/o \emph{warmup}, and (c)  the physics solver. The values inside the bars are the number of iterations it took for that stage. The number of \emph{warmup} iterations is indicated in parenthesis. At the top of each column is the speedup with respect to the physics solver.}
\label{fig:times}
\end{figure*}

The time to solution of CFDNet is the sum of the \emph{warmup}, \emph{inference}, and \emph{refinement} times which are defined as follows.

\emph{Warmup time, $t_{warmup}$} is the time for the physics solver to drop the error from the initial condition one order of magnitude (\ie $K$ iterations).
The physics solver is run in parallel as described in Section~\ref{sec:openfoam}.

\emph{Inference time, $t_{infer}$} is the sum of the times spent on each of the
following steps: (1) \texttt{MPI\_Gather} on the master process to get all the flow variables values from each MPI process
after warmup, (2) construction of the input tensor image, (3) CNN inference of
the output (the inference is computed on the CPU whose specifications are detailed in Section~\ref{sec:openfoam}), (4) \texttt{MPI\_Scatter} to distribute the output tensor image back to the MPI processes.

\emph{Refinement time, $t_{refine}$} is the time for the physics solver to further drop the error from the new initial conditions (\ie the output tensor after inference) four orders of magnitude for all variables in test sets B and C. For test set A, we drop 5 orders of magnitude for the velocity and pressure fields and 4 orders of magnitude for the eddy viscosity.
Note that different convergence criteria are used for each test set to correspond to the respective residuals (tolerances) used to create the training sets.
\emph{Refinement}, as the \emph{warmup}, is run in parallel.

Figure \ref{fig:times} compares the CFDNet time (broken down by $t_{warmup}$, $t_{infer}$, and $t_{refine}$) to the time the traditional physics solver implementation takes to drop the residual to the same user-defined tolerance defined above for each dataset. 
The values inside the bars are the number of iterations required by the physics solver in each stage (\ie $K$ for warmup, $M$ for inference, and $N$ for the physics-only solver).
We first focus on the end-to-end CFDNet performance and then, evaluate specifically, the importance of warmup in embedding domain-knowledge.

\textbf{\emph{CFDNet vs physics solver.}}
The leftmost stacked bar (a) for each test case represents the time taken by CFDNet and the rightmost bar (c) is the time taken by the OpenFOAM physics solver to meet the convergence criteria.
Across the board, CFDNet accelerates the simulations by a factor of $1.9 - 7.4\times$.
The highest performance gain is achieved for the OG-DF test cases followed by SG-SF and DG-SF last.
Compared to the physics solver-only simulation which requires 1578 iterations (1588 for extrapolated flow conditions) to drop the residual to \num{1e-5} for OG-DF, CFDNet only takes one CNN inference and 319 iterations (426 for extrapolated case) to reach the same convergence constraints.
SG-SF and DG-SF evaluate the framework's ability to generalize on unseen geometries (where the former uses a subset geometry while the latter has edges not included in the training datasets).
We observe that CFDNet outperforms the traditional physics solver on both use cases requiring fewer total number of iterations to drop the residual to meet the convergence constraints.

Recall from earlier that in contrast to the OG-DF channel flow case, the standalone network has difficulty inferring the right flow field around the airfoil and cylinder in the DG-SF flow cases. 
This is not surprising since prediction on a different geometry (whether it is inter- or extrapolated) is much more challenging than when the geometry is kept fixed.
However, it is interesting to note that the network finds the cylinder geometry particularly challenging to accelerate compared to the flow around the airfoil and interpolated ellipse.
Even though a cylinder is simply a special case of an ellipse from a geometrical perspective, the physics in the rear part of the cylinder is more non-linear (a large recirculation area) and therefore, more refinement iterations than the other cases are required. 
Nevertheless, CFDNet still outperforms the physics solver by $1.9 - 2.6\times$ for the DG-SF cylinder case.
These results show CFDNet's potential to generalize to geometries whose fluid flows are significantly different from those in the training set while outperforming state-of-the-art physics solvers.

\textbf{\emph{Importance of warmup.}}
We now compare the results presented in the previous section when the inference is done without \emph{domain-specific} knowledge (no \emph{warmup}). The (b) bars in Figure~\ref{fig:times} show the time taken by CFDNet when the warmup method is not applied. Without warmup, the speedups drop significantly to $1.2 - 2.3\times$. 

In CFD simulations, the \emph{domain-specific} knowledge is not embedded until the numerical scheme interacts with the values in the boundary conditions. For example, the user-given initial condition is domain-blind. This is the objective of \emph{warmup}: let the solver carry ($K$) initial iterations so that \emph{domain-specific} knowledge (\eg geometry definition, flow condition, etc) is embedded in the input tensor. 

In the flow around solid bodies case, having the geometry entirely defined in the input tensor is critical for the network to predict the right flow regime. This is specially true for the DG-SF turbulent cylinder. Without warmup, the network predicts a field quantitatively\footnote{Our computed RME in this case are 110\% for the velocity, 251\% for the pressure, and 80\% for the modified eddy viscosity.} poor compared to the physics solver solution. The \emph{iterative refinement} uses a lot of iterations to meet the physics solver convergence constraints. For this reason, we even observe a slowdown compared to the physics solver without warmup. 
To summarize, across all the flow configurations tested in this paper, warmup only takes $1-2\%$ of the overall iterations of the physics solver while yielding significant speedups of $1.9 - 4.6\times$.


\section{Related Work}
\label{sec:related}
\textbf{\emph{Computational Fluid Dynamics.}} Direct Numerical Simulation (DNS) is an attempt to solve the discretized Navier-Stokes equations accounting for all time and length scales of turbulence \cite{pope}.
This leads to very fine meshes and small timesteps which make DNS computationally intractable for several flows. 
Reynolds Averaged Navier-Stokes (RANS) equations \cite{reynolds} time-average the effects of turbulence at the expense of yielding a non-closed equation. Closing the equation has been extensively done through modeling \cite{SpalartAllmaras}. 
However, even with RANS coupled with turbulence modeling, exploring different design alternatives for complex flows and geometries is computationally expensive. 
To circumvent this, it is common to use surrogates - imperfect but computationally inexpensive models to accelerate design space exploration for industrial applications.

\textbf{\emph{Accelerating Fluid Simulations with Surrogates.}} Shape design optimization relies on the minimization of an objective function. 
Gaussian Processes (GP) approximate this objective function to be computationally tractable. 
However, these models rely on a specific geometry parametrization \cite{GP} which limits its generalizability. 
Other regression techniques, such as MARS and polynomial regressions \cite{MARS,polyRegressionA,polyRegressionB} only predict the velocity or pressure fields on particular points on the surface of the solid body, in order to reduce the dimensional complexity of the problem. 
In this paper, we take an alternate approach with neural networks that predicts the velocity and the pressure fields for all points in the flow field.

\textbf{\emph{Accelerating Fluid Simulations with Neural Networks.}} In the past few years, several researchers have leveraged neural networks to accelerate fluid dynamics simulations. 
\citeauthor{tompson} accelerate
Eulerian fluid simulations by replacing the Poisson solver step in an Eulerian
flow iterative solver instead of finding an end-to-end mapping to calculate
the divergence-free velocity field \cite{tompson}.
Alternatively, \citeauthor{autodesk} find a real-time solution to viscous laminar flows around
solid objects \cite{autodesk}. 
However, the above approaches have elemental constraints. 
First, Eulerian fluid
simulations ignore second-order velocity derivatives in the Navier-Stokes equations. 
Second, viscous laminar flow approaches are ambiguous if they are to be expanded to turbulent flows which are intrinsically chaotic and much harder to resolve \cite{pope}.

There have been recent attempts to find neural network-based accelerators
for turbulent flows. 
The results are promising but the neural networks predict only a subset of the flow variables. 
\citeauthor {maulik} predict the eddy viscosity field and not other flow properties such as velocity and pressure fields  \cite{maulik}.
\citeauthor{unet} use a novel input-output representation but their approach does not account for the eddy viscosity field \cite{unet}.
More importantly, the network prediction is limited to the fluid domain closest to the solid body, so it remains an unknown how the network would perform in the freestream, where the boundary conditions are set which define the flow configuration.
Another limitation of \cite{unet} is that the
network is used as a final, end-to-end surrogate. 
Even though the
results reported are promising (RME less than 3\%) and the
surrogate approach provides real-time solutions, the geometries in the training
and prediction stages are the same (airfoils). 
It remains unclear how the
surrogate would perform on different geometries.
CFDNet injects the neural network-based mapping back into the
physical solver to enable both extrapolation and generalization with the same model without relaxing the convergence constraints. 


\section{Conclusions}
\label{sec:conclusions}

This paper has shown that coupling of a RANS fluid flow simulation
with a convolutional neural network, an approach called CFDNet, can
significantly accelerate the convergence of the overall scheme.
CFDNet speeds the simulations up by a factor of
$1.9 - 7.4\times$ on both steady laminar and turbulent flows on a
variety of geometries, without relaxing the convergence constraints of
the physics solver. 
To evaluate the model's capacity for generalization and extrapolation,
it is tested across a range of scenarios and geometries, including
channel flow, ellipses, airfoils, and cylinders. In general, the model
performs well and demonstrates a capacity to make accurate
predictions even for geometries unseen during training. 

CFDNet indicates that coupling physical models with
data-driven machine learning models is a promising approach for
accelerating the convergence of simulations. However, more work
remains for this method to be a widely used tool for predictive
engineering. The explorations shown in this paper are focused on
fluid-flow simulations, in particular, RANS models, but the
fundamental approach is not expected to be limited to a particular
model or physical domain. For example, it is expected that CFDNet
could be successfully applied, with minimal effort, to Large Eddy
Simulations (LES). In principal, any discretized field of 
inputs should be amenable to the approach outlined in this paper, and
so a wide variety of domains in scientific computing, such as
molecular dynamics, material science, etc., could be
considered. This approach may not be limited to physical simulations,
and may be useful as a preconditioner or as a replacement for
algebraic multigrid or other numerical linear algebra
solvers. Regardless, the generalizability of the network
architecture remains an important open question. Many CNN networks
have demonstrated efficacy across a variety of related tasks with only
minimal re-training necessary. The capability of transfer learning or
at least leveraging a similar network architecture for a wide range of
tasks is an important future exploration. 


Finally, while the results in this paper maintained the accuracy of the
underlying simulation, a user of CFDNet will not desire to run a
traditional simulation in parallel to check the simulation results.
Thus, more research should be performed on quantifying the uncertainty
and robustness of coupling DL and traditional physical simulations.

\section{Acknowledgment}
This work was partly supported by the National Science Foundation (NSF) under the award number 1750549. Any opinions, findings and conclusions expressed in this material are those of the authors and do not necessarily reflect those of NSF. We thank Dr. Ferran Martí, Senior Aerodynamicist at Boeing, for his help in sectioning and defining iterative time-marching numerical schemes and in the design of the airfoil's computational grid. ©2020 Advanced Micro Devices, Inc.  All rights reserved. 
AMD, the AMD Arrow logo, and combinations thereof are trademarks of Advanced Micro Devices, Inc.  Other product names used in this publication are for identification purposes only and may be trademarks of their respective companies.

\bibliographystyle{ACM-Reference-Format}
\balance
\bibliography{mybib,distdl,FasterLearning,extra,vishnu,apoptosis,agd,mathstat}

\end{document}